\documentclass[showpacs,preprintnumbers,prb]{revtex4}
\usepackage{amsmath}
\usepackage{amssymb}
\usepackage{graphicx}
\usepackage{dcolumn}
\usepackage{bm}

\input{tcilatex}

\begin{document}

\title{Entry of Magnetic Flux into a Magnetically Shielded Type-II
Superconductor Filament}
\author{S.~V.~Yampolskii}
\altaffiliation[On leave from ]{Donetsk Institute for Physics and Technology, National Academy of Sciences
of Ukraine, 83114 Donetsk, Ukraine. Electronic address:
yampolsk@tgm.tu-darmstadt.de}
\author{Yu.~A.~Genenko}
\affiliation{Institut f\"{u}r Materialwissenschaft, Technische Universit\"{a}t Darmstadt,
D-64287 Darmstadt, Germany }
\date{\today}

\begin{abstract}
In the framework of the London approximation the magnetic flux penetration
into a type-II superconductor filament surrounded by a soft-magnet sheath
and exposed to a transverse external magnetic field is studied. The lower
transverse critical field as well as the critical field and the critical
current of the first vortex nucleation at the superconductor/magnet
interface are calculated on the basis of an exact solution for a vortex of
arbitrary plane configuration. The Bean-Livingston barrier against the
vortex nucleation is shown to strongly depend on the magnet sheath
parameters.
\end{abstract}

\maketitle

\section{Introduction}

Hybrid systems composed of magnetic and superconducting materials
attract much attention during last years in view of possibilities
to improve
superconductor critical parameters. There were conducted many experimental~%
\cite%
{Schuller1,Schuller2,Moshch1,Moshch1a,Moshch2,Moshch2a,Moshch3,Johansen2}
and theoretical~\cite%
{Peeters0,Pokrovsky1,Cheng,Bulaevsky,Pokrovsky2,Pokrovsky2a,Johansen,Peeters,Peeters2,Sonin,Helseth2,Melnikov,Erdin}
studies of the heterostructures composed of superconductors (SC) and \textit{%
ferromagnets} (FM). Diverse vortex configurations are generated in
such structures due to the large intrinsic magnetic moments of the
FM elements
(magnetic dots~\cite%
{Schuller2,Peeters0,Pokrovsky1,Pokrovsky2a,Peeters,Peeters2} or
inhomogeneities of the magnet
layer~\cite{Pokrovsky2,Johansen,Sonin}) and various transitions
between them occur. The interaction of vortices with
these intrinsic moments results in matching effects of a vortex lattice~\cite%
{Schuller1,Moshch1,Moshch3}, spontaneous nucleation of vortices
inside a superconductor
layer~\cite{Peeters,Peeters2,Melnikov,Erdin}, enhancement of
vortex pinning~\cite{Bulaevsky,Moshch1,Moshch1a,Moshch2,Moshch2a}
and
increase of the critical magnetic field of a superconductor~\cite%
{Moshch3,Cheng,Peeters2}.

Much less attention has been attracted by heterostructures of
superconductors and \textit{soft magnets} (SM). Soft magnets, such
as permalloy, pure iron, crioperm, etc., have, as a rule,
sufficiently large values of the relative permeability~$\mu $,
very narrow hysteresis loop and possess negligible remanent
magnetization. Nevertheless, they may significantly improve
superconductor performance by effective shielding from
the external magnetic field as well as from the transport current self-field~%
\cite{Genenko1,Genenko2,Campbell,Campbell2,Jooss,GRS1,GRS2,YGR}.
It was shown first theoretically~\cite{Genenko1,Genenko2} that the
magnetic shielding may increase the critical current of a
superconductor strip enhancing in this way its current-carrying
capability both in the Meissner and in the partly flux-filled
states. It was found also that such shielding can strongly reduce
the transport AC losses in superconductor wires and
tapes~\cite{Campbell,Campbell2}. Strong current redistribution in
superconductor strips due to bulk SM environments has been
established recently by magneto-optics~\cite{Jooss}.

After the discovery of superconductivity in magnesium
diboride~\cite{MgB2} very intense investigations were carried out
on superconducting MgB$_{2}$ wires sheathed in iron, which became
ideal objects to explore the magnetic
shielding effect due to simplicity of their fabrication~\cite%
{Dou1,Dousamples}. As was observed in recent experiments, such
structures exhibit enhanced superconducting critical currents over
a wide range of the external magnetic
field~\cite{Dou1,Dou2,Dou3,Dou4} as well as strong reduction of AC
losses in the external field~\cite{DouAC1,DouAC2,DouAC3}. At the
same time, a theoretical description of the influence of
soft-magnet shielding on the current-carrying properties of
type-II superconductor filaments is still lacking.

Recently, we have considered the flux-free Meissner state in a
type-II superconductor filament surrounded by a circular
soft-magnet sheath~\cite{VM} and calculated the field distribution
and the magnetic moment in this object. In the present paper we
consider properties of the vortex state in the above sample when
it is exposed to a transverse magnetic field and/or carries a
transport current. In particular, we derive general expressions
for the magnetic field of an arbitrary plane vortex and found the
transverse lower critical field~$H_{c1}$ and the field of the
first flux entry~$H_{p}$. In addition to the experimental
significance~\cite{Dou1,
Dousamples,Dou2,Dou3,Dou4,DouAC1,DouAC2,DouAC3} of such SC/SM
heterostructures, the system under consideration is simpler, from
the theoretical point of view, than the strip geometry, because it
allows one to exclude the strong influence of large geometrical
factor on the
superconducting response typical of the planar configurations~\cite%
{Genenko2,Genenko3,Brandt1,Brandt2}. In our consideration we will
follow conceptually Ref.~\cite{Genenko4} where the lower critical
field and the critical conditions for the first flux entry in a
current-carrying type-II superconducting cylinder exposed to a
transverse magnetic field have been established.

The paper is organized as follows. The theoretical model is
presented in Sec.~II. In Sec.~III we derive the magnetic field
distribution for a single vortex of an arbitrary plane shape and
give a general expression for the self-energy of the vortex in a
composite SC/SM filament. The dependences of vortex magnetic
moment on the thickness and the relative permeability of the
magnet sheath are discussed in Sec.~IV. Further we found critical
parameters of the SC/SM cylinder: the lower critical
field~$H_{c1}$ is obtained in Sec.~V and the conditions for the
vortex loop nucleation on the SC/SM interface are established in
Sec.~VI. Our conclusions are presented in Sec.~VII.

\section{Theoretical model}

\begin{figure}[h]
\includegraphics{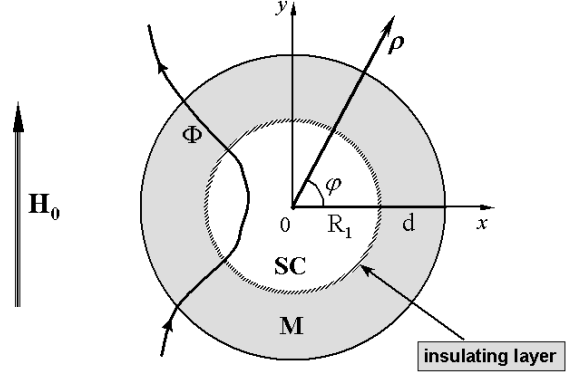}
\caption{Cross-sectional view of a superconductor filament covered by a
coaxial cylindrical magnetic sheath and exposed to external transverse
magnetic field. A plane single vortex of an arbitrary form entering a
superconductor is shown.}
\label{fig1}
\end{figure}

Let us consider an infinite type-II superconductor cylinder of
radius $R_{1}$ enveloped in a coaxial magnetic sheath of thickness
$d$ with a relative magnetic permeability $\mu >1$; the structure\
extended along the $z$\ axis of cylindrical coordinate system
$\left( \rho ,\varphi ,z\right) $\ adapted
to the cylinder (Fig.~\ref{fig1}). A transverse magnetic field $\mathbf{H}%
_{0}$ is applied along the positive $y$\ direction and is
asymptotically uniform at distances large compared to
$R_{2}=R_{1}+d$. In our consideration we will neglect the remanent
magnetization as well as both nonlinear behavior and conductivity
of the magnetic layer so that the magnetic induction
$\mathbf{B}=\mu _{0}\mu \mathbf{H}$ in the magnet and, therefore,
a relative permeability $\mu $ is assumed the only characteristic
of a homogeneous, isotropic SM sheath ($\mu _{0}$ is the
permeability of free space).

We start from the London equation for the magnetic induction $\mathbf{B}%
^{\left( 1\right) }$ in the superconductor area~\cite{deGennes}
\begin{subequations}
\begin{equation}
\mathbf{B}^{\left( 1\right) }+\lambda ^{2}\text{curl curl }\mathbf{B}%
^{\left( 1\right) }=\mathbf{\Phi },\quad \rho \leq R_{1},  \tag{1}
\label{eq:1}
\end{equation}

\noindent with the London penetration depth, $\lambda $, and the
source function describing an arbitrary vortex
\end{subequations}
\begin{subequations}
\begin{equation}
\mathbf{\Phi (r)}=\Phi _{0}\int d\mathbf{l}\,\delta
(\mathbf{r}-\mathbf{l}), \tag{2}  \label{eq:2}
\end{equation}

\noindent where $\Phi _{0}$ is the unit flux quantum, $\mathbf{r}$
is the position vector, $d\mathbf{l}$ is the flux-line element;
the integration extends along the flux-line (vortex core). The
magnetic field outside the superconductor denoted by
$\mathbf{H}^{\left( 2\right) }$\ in the magnetic sheath and by
$\mathbf{H}^{\left( 3\right) }$\ in the surrounding free space is
described by Maxwell equations
\end{subequations}
\begin{subequations}
\begin{equation}
\text{curl }\mathbf{H}=0,\quad \rho \geq R_{1},  \tag{3}
\label{eq:3}
\end{equation}

\end{subequations}
\begin{subequations}
\begin{equation}
\text{div }\mathbf{B}=0,  \tag{4}  \label{eq:4}
\end{equation}

\noindent the latter of which is valid in the whole space.

We imply the existence of an insulating, nonmagnetic layer of
thickness much less than $\lambda $, $d$\ and $R_{1}$\ between the
superconductor and the
magnet sheath (for example, such a layer was experimentally observed in MgB$%
_{2}/$Fe wires~\cite{Dou4}). According to this assumption Eqs. (\ref{eq:1}-%
\ref{eq:4}) are provided with the following boundary conditions
\end{subequations}
\begin{subequations}
\begin{eqnarray}
B_{n}^{\left( 1\right) } &=&\mu _{0}\mu H_{n}^{\left( 2\right)
},\quad B_{t}^{\left( 1\right) }=\mu _{0}H_{t}^{\left( 2\right) };
\TCItag{5a} \label{eq:5a} \\ \mu H_{n}^{\left( 2\right) }
&=&H_{n}^{\left( 3\right) },\quad H_{t}^{\left( 2\right)
}=H_{t}^{\left( 3\right) },  \TCItag{5b}  \label{eq:5b}
\end{eqnarray}%
for the normal ($n$) and tangential ($t$) field components on the
superconductor/magnet interface~(5a) and on the outer magnet
surface~(5b), respectively. In addition, the field has to approach
asymptotically the value of the external magnetic field
$\mathbf{H}_{0}$.

The solution of Eqs.~(\ref{eq:1}-\ref{eq:4}) may be represented as
a
superposition of the Meissner response $\mathbf{B}_{M}$, induced by $\mathbf{%
H}_{0}$ in absence of the magnetic vortex, and of the induction $\mathbf{b}$%
\ of the vortex itself. The field $\mathbf{B}_{M}$\ satisfies Eqs.~(\ref%
{eq:1}-5) with $\mathbf{\Phi }=0$\ and has been found
recently~\cite{VM}. Therefore we may rewrite Eq.~(\ref{eq:1}) in
the form
\end{subequations}
\begin{subequations}
\begin{equation}
\mathbf{b}^{\left( 1\right) }+\lambda ^{2}\text{curl curl }\mathbf{b}%
^{\left( 1\right) }=\mathbf{\Phi },\quad \rho \leq R_{1}.  \tag{6}
\label{eq:6}
\end{equation}

\noindent Taking into account that the field of\ the vortex is
potential
outside the superconductor and may be presented as $\mathbf{h}=-\mathbf{%
\nabla }\psi $\ we obtain in this area, instead of Eqs.~(\ref{eq:3})-(\ref%
{eq:4}), the Laplace equation \
\end{subequations}
\begin{subequations}
\begin{equation}
\Delta \psi =0,\quad \rho \geq R_{1},  \tag{7}  \label{eq:7}
\end{equation}

\noindent with $\psi \rightarrow 0$\ at $\rho \rightarrow \infty $
and the boundary conditions~(5) applying mutatis mutandis as well.

\section{Structure of a plane magnetic vortex in a SC/SM cylindrical sample}

In the manner of Ref.~\cite{Genenko4}, we look for the components
of the vortex self-field in cylindrical coordinates
$\mathbf{b}^{\left( 1\right) }=\left( b_{\rho }^{\left( 1\right)
},b_{\varphi }^{\left( 1\right) },b_{z}^{\left( 1\right) }\right)
$\ and for the potential $\psi $\ using the Fourier transformation
in the form
\end{subequations}
\begin{subequations}
\begin{equation}
b^{j\left( 1\right) }\left( \rho ,\varphi ,z\right)
=\sum\limits_{m}\exp \left( im\varphi \right) \int \frac{dk}{2\pi
}b_{k,m}^{j\left( 1\right) }\left( \rho \right) \exp \left(
-ikz\right) ,  \tag{8}  \label{eq:8}
\end{equation}

\end{subequations}
\begin{subequations}
\begin{equation}
\psi \left( \rho ,\varphi ,z\right) =\sum\limits_{m}\exp \left(
im\varphi \right) \int \frac{dk}{2\pi }\psi _{k,m}\left( \rho
\right) \exp \left( -ikz\right) ,  \tag{9}  \label{eq:9}
\end{equation}

\noindent where the index $j$\ assumes values $\rho ,\varphi ,z$.
In terms of the Fourier amplitudes $b_{k,m}^{j\left( 1\right) }$\
and $\psi _{k,m}$\ Eqs.(\ref{eq:6})-(\ref{eq:7}) transform to the
set of ordinary differential equations
\end{subequations}
\begin{subequations}
\begin{equation}
\frac{\partial ^{2}b_{k,m}^{\rho \left( 1\right) }}{\partial \rho ^{2}}+%
\frac{1}{\rho }\frac{\partial b_{k,m}^{\rho \left( 1\right)
}}{\partial \rho }-\left( Q^{2}+\frac{m^{2}+1}{\rho ^{2}}\right)
b_{k,m}^{\rho \left(
1\right) }-\frac{2im}{\rho ^{2}}b_{k,m}^{\varphi \left( 1\right) }=-\frac{%
\Phi _{k,m}^{\rho }}{\lambda ^{2}},\quad \rho \leq R_{1},
\tag{10} \label{eq:10}
\end{equation}

\end{subequations}
\begin{subequations}
\begin{equation}
\frac{\partial ^{2}b_{k,m}^{\varphi \left( 1\right) }}{\partial \rho ^{2}}+%
\frac{1}{\rho }\frac{\partial b_{k,m}^{\varphi \left( 1\right)
}}{\partial \rho }-\left( Q^{2}+\frac{m^{2}+1}{\rho ^{2}}\right)
b_{k,m}^{\varphi \left( 1\right) }+\frac{2im}{\rho
^{2}}b_{k,m}^{\rho \left( 1\right) }=-\frac{\Phi _{k,m}^{\varphi
}}{\lambda ^{2}},\quad \rho \leq R_{1},  \tag{11} \label{eq:11}
\end{equation}

\end{subequations}
\begin{subequations}
\begin{equation}
\frac{\partial ^{2}b_{k,m}^{z\left( 1\right) }}{\partial \rho ^{2}}+\frac{1}{%
\rho }\frac{\partial b_{k,m}^{z\left( 1\right) }}{\partial \rho
}-\left( Q^{2}+\frac{m^{2}}{\rho ^{2}}\right) b_{k,m}^{z\left(
1\right) }=-\frac{\Phi _{k,m}^{z}}{\lambda ^{2}},\quad \rho \leq
R_{1},  \tag{12}  \label{eq:12}
\end{equation}

\end{subequations}
\begin{subequations}
\begin{equation}
\frac{\partial ^{2}\psi _{k,m}}{\partial \rho ^{2}}+\frac{1}{\rho }\frac{%
\partial \psi _{k,m}}{\partial \rho }-\left( k^{2}+\frac{m^{2}}{\rho ^{2}}%
\right) \psi _{k,m}=0,\quad \rho \geq R_{1},  \tag{13}
\label{eq:13}
\end{equation}

\noindent with the boundary conditions
\end{subequations}
\begin{align}
b_{k,m}^{\rho \left( 1\right) }\left( R_{1}\right) & =-\mu _{0}\mu \frac{%
\partial \psi _{k,m}^{\left( 2\right) }\left( R_{1}\right) }{\partial \rho }%
,\quad ib_{k,m}^{\varphi \left( 1\right) }\left( R_{1}\right) =\mu _{0}\frac{%
m}{R}\psi _{k,m}^{\left( 2\right) }\left( R_{1}\right) ,\quad
b_{k,m}^{z\left( 1\right) }\left( R_{1}\right) =ik\mu _{0}\psi
_{k,m}^{\left( 2\right) }\left( R_{1}\right) ,  \tag{14}
\label{BC2} \\
\mu \frac{\partial \psi _{k,m}^{\left( 2\right) }\left( R_{2}\right) }{%
\partial \rho }& =\frac{\partial \psi _{k,m}^{\left( 3\right) }\left(
R_{2}\right) }{\partial \rho },\quad \psi _{k,m}^{\left( 2\right)
}\left( R_{2}\right) =\psi _{k,m}^{\left( 3\right) }\left(
R_{2}\right) ,\quad \psi _{k,m}^{\left( 3\right) }\left( \infty
\right) =0.  \notag
\end{align}

\noindent Here the value $Q=\left( k^{2}+\lambda ^{-2}\right)
^{1/2}$\ is introduced and the Fourier amplitudes $\Phi
_{k,m}^{j}$\ of the source function~(\ref{eq:2})\ are defined in
the same manner as the field components in Eq.~(\ref{eq:8}).
Indices (2) and~(3) in Eqs.~(\ref{BC2}) correspond to the areas of
the magnetic sheath and the surrounding free space, respectively.
Notice that Eqs.~(\ref{eq:10})-(\ref{eq:12}) are not identical to
Eq.~(\ref{eq:6}) since by their derivation we used the equality
curl~curl~$\mathbf{b}^{\left( 1\right) }=-\Delta
\mathbf{b}^{\left( 1\right) }$\ which implies that
div~$\mathbf{b}^{\left( 1\right) }=0$. Therefore the solutions of
Eqs.~(\ref{eq:10})-(\ref{eq:13}) should also satisfy the latter
constraint to be the solutions of Eqs.(\ref{eq:6})-(\ref{eq:7}).

We consider below arbitrary configurations of a plane vortex lying
in the
plane $z=0$, so that $\Phi _{k,m}^{z}=0$. Upon the transformation $%
f_{k,m}^{\pm }=b_{k,m}^{\rho \left( 1\right) }\pm
ib_{k,m}^{\varphi \left( 1\right) }$\ the set of
Eqs.~(\ref{eq:10})-(\ref{eq:13}) may be decoupled and solved it
terms of the modified Bessel functions. We obtain the solutions
regular at $\rho =0$:
\begin{align}
\binom{b_{k,m}^{\rho \left( 1\right) }}{ib_{k,m}^{\varphi \left( 1\right) }}%
& =\frac{1}{2}\left\{ I_{m+1}\left( Q\rho \right) \left[ F_{k,m}^{+}-\int%
\limits_{\rho }^{R_{1}}d\rho ^{\prime }\rho ^{\prime }\eta
_{k,m}^{+}\left( \rho ^{\prime }\right) K_{m+1}\left( Q\rho
^{\prime }\right) \right] \right. \tag{15}  \label{eq:15} \\ & \pm
I_{m-1}\left( Q\rho \right) \left[ F_{k,m}^{-}-\int\limits_{\rho
}^{R_{1}}d\rho ^{\prime }\rho ^{\prime }\eta _{k,m}^{-}\left( \rho
^{\prime }\right) K_{m-1}\left( Q\rho ^{\prime }\right) \right]
\notag \\ & \left. -K_{m+1}\left( Q\rho \right)
\int\limits_{0}^{\rho }d\rho ^{\prime }\rho ^{\prime }\eta
_{k,m}^{+}\left( \rho ^{\prime }\right) I_{m+1}\left( Q\rho
^{\prime }\right) \mp K_{m-1}\left( Q\rho \right)
\int\limits_{0}^{\rho }d\rho ^{\prime }\rho ^{\prime }\eta
_{k,m}^{-}\left( \rho ^{\prime }\right) I_{m-1}\left( Q\rho
^{\prime }\right) \right\} , \notag
\end{align}

\begin{subequations}
\begin{equation}
b_{k,m}^{z\left( 1\right) }=C_{k,m}I_{m}\left( Q\rho \right) ,
\tag{16} \label{eq:16}
\end{equation}

\end{subequations}
\begin{subequations}
\begin{equation}
\psi _{k,m}^{(2)}=\alpha _{k,m}I_{m}\left( \left\vert k\right\vert
\rho \right) +\beta _{k,m}K_{m}\left( \left\vert k\right\vert \rho
\right) , \tag{17}  \label{eq:17}
\end{equation}

\end{subequations}
\begin{subequations}
\begin{equation}
\psi _{k,m}^{(3)}=\Psi _{k,m}K_{m}\left( \left\vert k\right\vert
\rho \right) ,  \tag{18}  \label{eq:18}
\end{equation}

\noindent where $\eta _{k,m}^{\pm }\left( \rho \right) =-\lambda
^{-2}\left(
\Phi _{k,m}^{\rho }\pm i\Phi _{k,m}^{\varphi }\right) ,$ $I_{\nu }$ and $%
K_{\nu }$\ are the modified Bessel functions~\cite{Abramovitz}.
The
coefficients in Eqs.~(\ref{eq:15})-(\ref{eq:18}) are found from Eqs.~(\ref%
{BC2}) and the constraint div~$\mathbf{b}^{\left( 1\right) }=0$\
and are given in Appendix~A.

Before calculation of the physical properties of vortices of
definite configurations and critical parameters of the system
under consideration, we write here a general formula for the free
energy of an arbitrary plane vortex in terms of the above
presented solution. The self-energy of the system containing a
single vortex takes a form (see Appendix~B)
\end{subequations}
\begin{align}
F& =\frac{1}{2\mu _{0}}\int\nolimits_{\rho \leq R_{1}}dV\left[ \mathbf{b}%
^{\left( 1\right) 2}+\lambda ^{2}\left( \text{curl
}\mathbf{b}^{\left( 1\right) }\right) ^{2}\right]  \tag{19}
\label{eq:19} \\ & +\frac{\mu _{0}\mu }{2}\int\nolimits_{R_{1}\leq
\rho \leq R_{2}}dV~\left(
\mathbf{\nabla }\psi ^{\left( 2\right) }\right) ^{2}+\frac{\mu _{0}}{2}%
\int\nolimits_{\rho \geq R_{2}}dV~\left( \mathbf{\nabla }\psi
^{\left( 3\right) }\right) ^{2}  \notag \\ & =\frac{1}{2\mu
_{0}}\int\nolimits_{\rho \leq R_{1}}dV~\mathbf{b}^{\left( 1\right)
}\mathbf{\Phi }+\frac{R_{1}}{2}\sum\limits_{m}\int dk~\psi
_{k,m}^{(2)}\left( R_{1}\right) \Phi _{-k,-m}^{\rho }\left(
R_{1}\right) . \notag
\end{align}

\noindent It is interesting to note that this expression is
identical to the corresponding formula for the self-energy of a
vortex in a non-shielded superconducting cylinder (see Eq.~(21) in
Ref.~\cite{Genenko4}); the presence of a magnet is accounted for
implicitly by the actual form of the potential $\psi ^{(2)}$ and
of the magnetic induction $\mathbf{b}^{\left( 1\right) }$.

The above described solution~(\ref{eq:15}-\ref{eq:18}) exhibits a
proper transformation to the case of an isolated
superconductor~\cite{Genenko4} by
setting $\mu =1$. Together with the expression for the vortex self-energy~(%
\ref{eq:19}) it may be applied to any single or multiple plane
vortex configurations. Among others it allows one to investigate,
by special choice of the vortex shape, the lower critical field
and the magnetic flux entry in shielded superconductors carrying a
transport current and/or subjected to an external magnetic field,
the problems considered below.

\section{Magnetic moment of an arbitrary plane vortex loop}

In this section we consider a magnetic moment of a magnetically
shielded wire, an important measurable characteristic which is
also necessary for evaluation of critical parameters of the SC/SM
heterostructure. The magnetic
moment projection on the field $\mathbf{H}_{0}$\ direction (see Fig.~\ref%
{fig1}) consists of two parts, presenting contributions from the
superconductor and from the magnetic sheath as follows
\begin{subequations}
\begin{equation}
M_{y}=M_{y}^{\left( 1\right) }+M_{y}^{\left( 2\right) }=\frac{1}{2}%
\dint\nolimits_{\rho \leq R_{1}}dV\left[ \mathbf{\rho }\times \mathbf{j}%
\right] _{y}-\left( \mu -1\right) \int\nolimits_{R_{1}\leq \rho
\leq R_{2}}dV~\left( \mathbf{\nabla }\psi ^{\left( 2\right)
}\right) _{y}, \tag{20}  \label{eq:20}
\end{equation}

\noindent which may be reduced to forms
\end{subequations}
\begin{subequations}
\begin{equation}
M_{y}^{\left( 1\right) }=-\frac{2\pi }{\mu
_{0}}R_{1}^{2}b_{0,1}^{\varphi
\left( 1\right) }\left( R_{1}\right) +\frac{2\pi i}{\mu _{0}}%
\dint\limits_{0}^{R_{1}}d\rho ~\rho \left( b_{0,1}^{\rho \left(
1\right) }-ib_{0,1}^{\varphi \left( 1\right) }\right) ,  \tag{21}
\label{eq:21}
\end{equation}

\end{subequations}
\begin{subequations}
\begin{equation}
M_{y}^{\left( 2\right) }=2\pi i\left( \mu -1\right) \left[
R_{2}\psi _{0,1}^{\left( 2\right) }\left( R_{2}\right) -R_{1}\psi
_{0,1}^{\left( 2\right) }\left( R_{1}\right) \right] .  \tag{22}
\label{eq:22}
\end{equation}

Let us consider now an arbitrary vortex loop lying in the plane
$z=0$\ and penetrating the superconductor to the depth of $r$ from
the surface. For simplicity we also suppose that the loop is
symmetric with respect to $x$\ axis and, therefore, we will
describe its specific form by some smooth
monotonic function $\varphi =\chi \left( \rho \right) $. Let us denote $%
R_{1}-r$ as the least value of radius $\rho $ for which the loop
exists, then $\chi \left( R_{1}-r\right) =0$. With function $\chi
\left( \rho \right) $\ defined in this way the Fourier amplitudes
of the source function~(\ref{eq:2}) read~\cite{Genenko4}:
\end{subequations}
\begin{align}
\Phi _{k,m}^{\rho }& =\frac{\Phi _{0}}{i\pi \rho }\sin \left[
m\chi \left( \rho \right) \right] \theta \left( \rho
-R_{1}+r\right) ,\quad \Phi _{k,m}^{z}=0,  \tag{23}  \label{eq:23}
\\ \Phi _{k,m}^{\varphi }& =\frac{\Phi _{0}}{\pi }\frac{d\chi
\left( \rho \right) }{d\rho }\cos \left[ m\chi \left( \rho \right)
\right] \theta \left( \rho -R_{1}+r\right) .  \notag
\end{align}

\noindent Then, upon the substitution of the amplitudes $\Phi
_{0,1}^{\rho }$
and $\Phi _{0,1}^{\varphi }$ in the general expressions (\ref{eq:15}), (\ref%
{eq:17}) and (\ref{eq:20}-\ref{eq:22}), we obtain the magnetic
moment of the loop:
\begin{equation}
M_{y}=M_{y}^{0}\frac{2\mu +\left( \mu ^{2}+1\right) \left(
d/R_{1}\right) \left( 2+d/R_{1}\right) }{2\mu +\left( \mu
+1\right) \left[ 1+\left( \mu -1\right) I_{1}^{\prime }\left(
R_{1}/\lambda \right) /I_{0}\left(
R_{1}/\lambda \right) \right] \left( d/R_{1}\right) \left( 2+d/R_{1}\right) }%
,  \tag{24}  \label{eq:24}
\end{equation}

\noindent where the prime denotes the derivative of the Bessel
function with respect to its argument, and
\begin{equation}
M_{y}^{0}=\frac{4\Phi _{0}}{\mu _{0}\lambda }\frac{1}{I_{0}\left(
R_{1}/\lambda \right) }\int\limits_{R_{1}-r}^{R_{1}}d\rho \rho
I_{1}\left( \frac{\rho }{\lambda }\right) \sin \left[ \chi \left(
\rho \right) \right] \tag{25}  \label{eq:25}
\end{equation}

\noindent is the magnetic moment for the unshielded superconductor filament~%
\cite{Genenko4} (notice that the formula (A4) in
Ref.~\cite{Genenko4} may be reduced to the above form). As it
follows from Eq.~(\ref{eq:24}), the dependence of the magnetic
moment on the relative permeability $\mu $ and the thickness $d$
of the magnet sheath is universal in the sense that it does not
depend on the specific form of the vortex. It is interesting to
note that the magnetic moment may be factorized as in
Eq.~(\ref{eq:24}) though the contributions of the superconductor
and of the magnet sheath are superimposed in the
definition~(\ref{eq:20}).

\begin{figure}[h]
\includegraphics{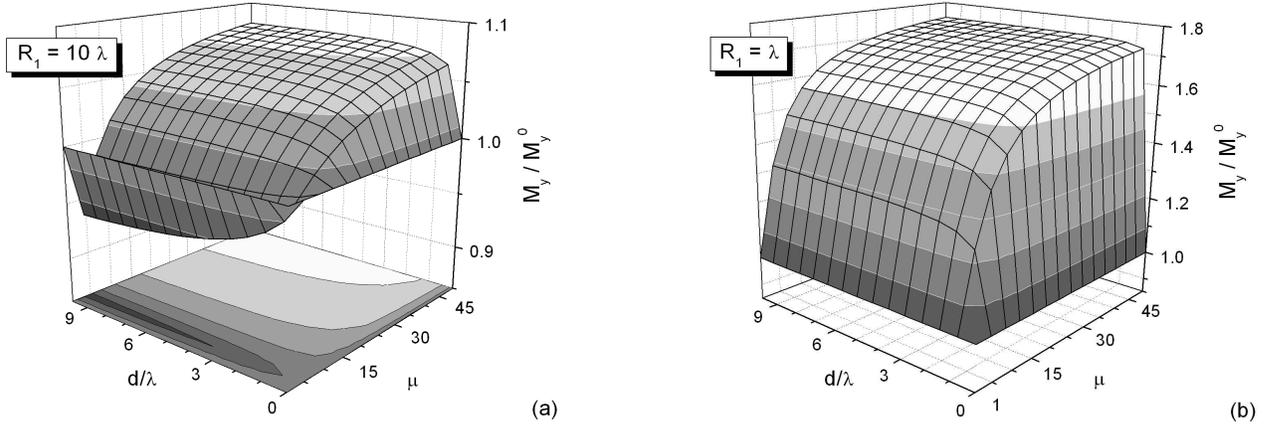}
\caption{The dependence of the vortex magnetic moment $M_{y}$ on the
relative permeability $\protect\mu $ and on the thickness $d$ of the magnet
sheath for different values of the superconductor radius: a)~$R_{1}=10%
\protect\lambda $, and b)~$R_{1}=\protect\lambda $.}
\label{fig2}
\end{figure}

The $M_{y}\left( \mu ,d\right) $ dependence is shown in
Fig.~\ref{fig2} for
two different values of superconductor radius: $R_{1}=10\lambda $ and $%
R_{1}=\lambda $. One can see that this dependence is different for
thick and
thin superconductors. For enough thick superconductor (see Fig.~\ref{fig2}%
,a) the moment $M_{y}$ reveals a minimum as a function of $\mu $\
for any fixed $d$. In the limit of $R_{1}\gg \lambda $\ the
$M_{y}\left( \mu ,d\right) $ dependence is described by the
expression
\begin{equation}
M_{y}=M_{y}^{0}\frac{2\mu +\left( \mu ^{2}+1\right) \left(
d/R_{1}\right) \left( 2+d/R_{1}\right) }{2\mu +\mu \left( \mu
+1\right) \left( d/R_{1}\right) \left( 2+d/R_{1}\right) }.
\tag{26}  \label{eq:26}
\end{equation}

\noindent The minimum value of the moment is reached at the
permeability
\begin{equation}
\mu _{\ast }=1+\left[ 2+\frac{2}{\left( d/R_{1}\right) \left(
2+d/R_{1}\right) }\right] ^{1/2}  \tag{27}  \label{eq:27}
\end{equation}

\noindent and equals
\begin{equation}
M_{y}^{\ast }=M_{y}^{0}\frac{\mu _{\ast }^{2}-1}{\mu _{\ast
}^{2}}.  \tag{28} \label{eq:28}
\end{equation}

\noindent In the case of very thick magnet sheath ($d\gg R_{1}$)
the lowest value of the moment is $M_{y}^{\ast }\simeq
0.83M_{y}^{0}$.

In the limit of thin superconductor $R_{1}\ll \lambda $ the magnetic moment $%
M_{y}$ monotonically increases with increase of $\mu $\ or $d$\ (see Fig.~%
\ref{fig2},b) and this dependence is described by the expression
\begin{equation}
M_{y}=M_{y}^{0}\frac{2\left[ 2\mu +\left( \mu ^{2}+1\right) \left(
d/R_{1}\right) \left( 2+d/R_{1}\right) \right] }{4\mu +\left( \mu
+1\right) ^{2}\left( d/R_{1}\right) \left( 2+d/R_{1}\right) }.
\tag{29}  \label{eq:29}
\end{equation}%
Let us note that in both limiting cases the factor accounting for
the magnet sheath does not depend on~$\lambda $.

For investigation of critical parameters of the SC/SM
heterostructure we must specify a form of the vortex which will be
done in the next sections.

\section{Lower critical field $H_{c1}$ of a SC/SM cylinder in a transverse
external field}

\begin{figure}[h]
\includegraphics{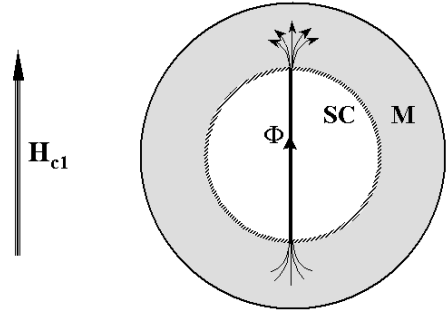}
\caption{Magnetic vortex lying along the diameter of a cylinder and parallel
to the external transverse magnetic field.}
\label{fig3}
\end{figure}

To obtain the value of the lower critical magnetic field we
consider now the case of a vortex taking a stable position in the
center of the sample,
namely directed along the cylinder diameter parallel to the applied field $%
\mathbf{H}_{0}$ (Fig.~\ref{fig3}).\ This position is analogous to
that of the straight vortex located deep inside a bulk
superconductor cylinder parallel to the external field which
energy defines the lower critical field for bulk
samples~\cite{deGennes}. The central location of the vortex
apparently leads to a local minimum of the Gibbs free energy of
the system
\begin{equation}
G=F-\mathbf{MH}_{0},  \tag{30}  \label{eq:30}
\end{equation}

\noindent where $F$\ is the self-energy~(\ref{eq:19}) and
$\mathbf{M}$\ is
the magnetic moment of the sample due to the presence of the vortex~(\ref%
{eq:24}). Vanishing of the energy~(\ref{eq:30}) defines the value
of the lower critical field $H_{c1}$ at which the vortex becomes
firstly energetically favorable deep inside the superconductor
(notice that, in this case, the Meissner contribution to the
energy is constant and may be omitted).

For the vortex lying along the cylinder diameter the Fourier
amplitudes of the source function~(\ref{eq:2})
read~\cite{Genenko4}:
\begin{align}
\Phi _{k,m}^{\rho }& =\frac{\Phi _{0}}{i\pi \rho }\sin \frac{\pi
m}{2},\quad \rho \leq R_{1},  \tag{31}  \label{eq:31} \\ \Phi
_{k,m}^{\varphi }& =\Phi _{k,m}^{z}=0.  \notag
\end{align}

Upon the substitution of Eq.~(\ref{eq:31}) into the solution~(\ref{eq:15}-%
\ref{eq:18}), one can find approximate expressions for the
self-energy of the vortex~(\ref{eq:19}) in two limiting cases:
\begin{equation}
F\simeq \frac{\Phi _{0}^{2}}{2\pi \mu _{0}\lambda ^{2}}R_{1}\left[ \ln \frac{%
4R_{1}}{e\xi }+\frac{6}{\pi }+O\left( R_{1}/\lambda \right)
\right] ,\qquad \xi \ll R_{1}\ll \lambda ,  \tag{32}
\label{eq:32}
\end{equation}%
\begin{equation}
F\simeq \frac{\Phi _{0}^{2}}{2\pi \mu _{0}\lambda ^{2}}R_{1}\left[ \ln \frac{%
\lambda }{\xi }-\gamma +O\left( \lambda /R_{1}\right) \right]
,\qquad \qquad \quad R_{1}\gg \lambda ,  \tag{33}  \label{eq:33}
\end{equation}

\noindent where the divergence of the energy at large
momentum~$k$, usual in the London theory~\cite{deGennes}, is cut
at the scale $k\sim 1/\xi $ with the superconductor coherence
length~$\xi $. One can see that in both cases of thin ($R_{1}\ll
\lambda $) and thick ($R_{1}\gg \lambda $) superconductors the
self-energy does not depend on the characteristics of magnet
sheath even in terms of the order of small parameters
$R_{1}/\lambda $ and, respectively, $\lambda /R_{1}$. The
dependence on $\mu $ may appear only in terms of the higher orders
of that small parameters. Therefore, the magnet sheath virtually
does not influence the self-energy of the vortex.

The magnetic moment of the sample is defined by Eq.~(\ref{eq:24}), where $%
M_{y}^{0}$ is easily obtained from Eq.~(\ref{eq:25}) with $\chi
\left( \rho \right) \equiv \pi /2$ (in this case the integration
over $\rho $\ starts from $0$):
\begin{equation}
M_{y}^{0}=\frac{2\pi \Phi _{0}R_{1}}{\mu _{0}}\left[
\frac{L_{0}\left( R_{1}/\lambda \right) }{I_{0}\left(
R_{1}/\lambda \right) }I_{1}\left( R_{1}/\lambda \right)
-L_{1}\left( R_{1}/\lambda \right) \right] ,  \tag{34}
\label{eq:34}
\end{equation}

\noindent where $L_{\nu }$\ is the modified Struve function.
Notice that, contrary to the self-energy, the magnetic moment
$M_{y}$\ strongly depends on $\mu $\ and $d$. Finally, from
Eq.~(\ref{eq:30}) we easily obtain the following expression for
the lower critical field:
\begin{equation}
H_{c1}=H_{c1}^{0}\frac{2\mu +\left( \mu +1\right) \left[ 1+\left(
\mu -1\right) I_{1}^{\prime }\left( R_{1}/\lambda \right)
/I_{0}\left( R_{1}/\lambda \right) \right] \left( d/R_{1}\right)
\left( 2+d/R_{1}\right) }{2\mu +\left( \mu ^{2}+1\right) \left(
d/R_{1}\right) \left( 2+d/R_{1}\right) },  \tag{35}  \label{eq:35}
\end{equation}

\noindent where $H_{c1}^{0}$ is the transverse lower critical
field of the non-sheathed sample, calculated in
Ref.~\cite{Genenko4}. Notice that by virtue of definition the
$H_{c1}\left( \mu ,d\right) $ dependence is inverted to that of
the magnetic moment~(\ref{eq:24}).

\begin{figure}[h]
\includegraphics{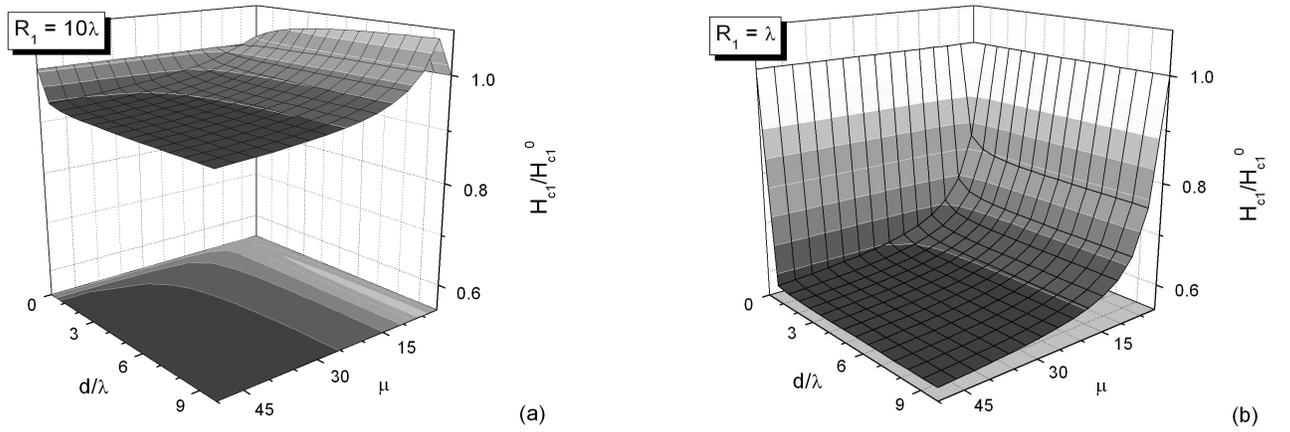}
\caption{The dependence of the lower critical field $H_{c1}$ on the relative
permeability $\protect\mu$ and on the thickness $d$ of the magnet sheath for
different values of the superconductor radius: a)~$R_{1}=10 \protect\lambda $%
, and b)~$R_{1}= \protect\lambda $.}
\label{fig4}
\end{figure}

The dependences of $H_{c1}$ on $\mu $ and $d$ are shown in
Fig.~\ref{fig4} for the same values of radius of the
superconductor as in Fig.~\ref{fig2}. One can see that for enough
large radii this dependence is nonmonotonic and reveals the region
of magnet permeability values where $H_{c1}>H_{c1}^{0}$. For
$R_{1}=10\lambda $ (Fig.~\ref{fig4},a) the lower critical field is
enhanced up to 10\% and this enhancement grows with increase of
$R_{1}$. In the limiting case $R_{1}\gg \lambda $ the
$H_{c1}\left( \mu ,d\right) $ dependence is described by the
expression
\begin{equation}
H_{c1}=H_{c1}^{0}\frac{\mu \left[ 2+\left( \mu +1\right) \left(
d/R_{1}\right) \left( 2+d/R_{1}\right) \right] }{2\mu +\left( \mu
^{2}+1\right) \left( d/R_{1}\right) \left( 2+d/R_{1}\right) }.
\tag{36} \label{eq:36}
\end{equation}

\noindent The maximum value of $H_{c1}$ is reached at any fixed
value of the magnet layer thickness $d$ for the permeability $\mu
_{\ast }$~(\ref{eq:27}) and equals
\begin{equation}
H_{c1}=H_{c1}^{0}\frac{\mu _{\ast }^{2}}{\mu _{\ast }^{2}-1},
\tag{37} \label{eq:37}
\end{equation}%
taking on the largest value $H_{c1}\cong 1.2H_{c1}^{0}$ at $d\gg
R_{1}$. With decrease of $R_{1}$ this effect disappears and for
$R_{1}\lesssim \lambda $\ the presence of magnet sheath depresses
the lower critical field (see Fig.~\ref{fig4},b). In the limit of
thin superconductor core $R_{1}\ll \lambda $ the critical field is
described by the asymptotic expression:
\begin{equation}
H_{c1}=\frac{H_{c1}^{0}}{2}\frac{4\mu +\left( \mu +1\right)
^{2}\left( d/R_{1}\right) \left( 2+d/R_{1}\right) }{2\mu +\left(
\mu ^{2}+1\right) \left( d/R_{1}\right) \left( 2+d/R_{1}\right) }.
\tag{38}  \label{eq:38}
\end{equation}

A practical conclusion here is that a cylindrical magnet sheath
has a detrimental effect on superconductivity in thin
superconductor wires of radius less than $\lambda $ facilitating
vortex phase at lower magnetic fields. On the other hand, the
magnetic coating of thick superconductors with radius much larger
than $\lambda $ allows optimization of the sheath parameters $d$
and $\mu $ in reasonable ranges leading to the moderate
enhancement of the lower critical field.

\section{Vortex loop nucleation at the SC/SM interface (the Bean-Livingston
barrier)}

\begin{figure}[h]
\includegraphics{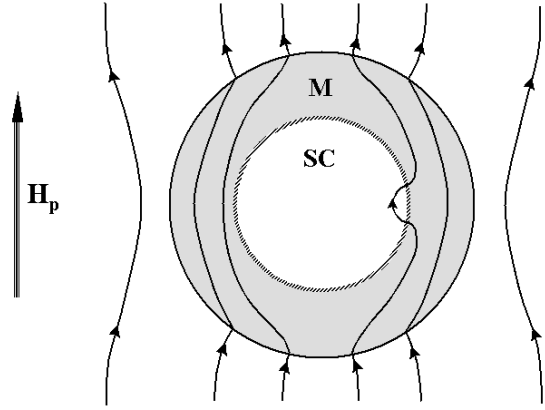}
\caption{Scheme of the first vortex loop nucleation at the interface between
superconductor and magnet sheath in the SC/SM filament exposed to an
external magnetic field.}
\label{fig5}
\end{figure}

In the non-shielded type-II superconductor sample exposed to a
transverse magnetic field the entry of magnetic flux starts with
the small loop nucleation at the sample
surface~\cite{Galajko,Petukhov,Koshelev,Samokhvalov} when the
surface Bean-Livingston barrier~\cite{Bean} is overcome. It is
evident that in the SC/SM system concerned the similar process of
vortex loop nucleation takes place at the interface between the
superconducting core and the magnet sheath (Fig.~\ref{fig5}).
However, due to the magnetization of magnetic medium, the
nucleation of vortex loop at the SC/SM interface may differ from
this on the uncovered SC cylinder surface studied
earlier~\cite{Genenko4}.

\subsection{Nucleation of vortex loop in a transverse magnetic field}

The Bean-Livingston barrier is a result of competition between the
attraction of the vortex to the boundary and the repulsive Lorentz
force exerted upon the vortex by the Meissner current. To evaluate
the critical field of the first vortex loop penetration into the
SC cylinder it is convenient to present the Gibbs energy of the
system as a sum of the vortex loop free energy and the work of the
external source of the magnetic field calculated as the work of
the Meissner current~\cite{Shmidt},
\begin{equation}
G=F-\Delta W_{H}.  \tag{39}  \label{eq:39}
\end{equation}

We consider a small semicircle loop of radius $a\ll \lambda $\
defined in Ref.~\cite{Genenko4} by the source
function~(\ref{eq:23}) with
\begin{equation}
\chi \left( \rho \right) =\phi _{t}\theta \left( R_{1}-\rho
\right) \theta \left( \rho -\sqrt{R_{1}^{2}-a^{2}}\right)
+\widetilde{\chi }\left( \rho
\right) \theta \left( \rho -R_{1}+a\right) \theta \left( \sqrt{%
R_{1}^{2}-a^{2}}-\rho \right) ,  \tag{40}  \label{eq:40}
\end{equation}

\noindent where $\sin \phi _{t}=a/R_{1}$\ and $\cos
\widetilde{\chi }\left( \rho \right) =\left( R_{1}^{2}+\rho
^{2}-a^{2}\right) /2R_{1}\rho $. Substituting the amplitudes $\Phi
_{k,m}^{j}$ into the general expression
for energy~(\ref{eq:19}) with magnetic field components from Eqs.~(\ref%
{eq:15})-(\ref{eq:18}) we find that the self-energy of the vortex
loop for
two limiting cases of thick $\left( R_{1}\gg \lambda \right) $ and thin $%
\left( R_{1}\ll \lambda \right) $ superconducting core coincides,
in the
main approximation, with the result for the unshielded SC cylinder~\cite%
{Genenko4},
\begin{equation}
F\simeq \frac{\Phi _{0}^{2}}{4\pi \mu _{0}\lambda ^{2}}\pi a\ln
\frac{a}{\xi },\qquad \xi \ll a\ll \lambda ,  \tag{41}
\label{eq:41}
\end{equation}

\noindent and does not depend on parameters of magnetic sheath as
well as in the case of the straight vortex along the diameter (see
Sec.~V). The dependence of the loop self-energy on permeability
$\mu $ appears only in higher orders of the small parameter
$a/\lambda $. Because of complexity of the general expressions
(\ref{eq:15})-(\ref{eq:18}) it is hard to exactly derive these
terms. Fortunately, it is sufficient here to estimate the
difference of the loop energy in the case under consideration from
the case of unshielded superconductor. This difference reaches its
maximum in the limit of an infinitely large~$\mu $ (see, for
example, Ref.~\cite{Genenko3}) and is less than the main
approximation~(\ref{eq:41}) by the factor of the order $a/\lambda
$. Therefore, we conclude that the self-energy of the vortex loop
is not affected substantially by presence of the magnet.

The above paradoxical result, that the influence of the magnet
sheath on the vortex self-energy is inessential even for large
permeabilities, can be explained in the following way. Apart from
the major contribution to the
vortex energy~(\ref{eq:41}) proportional to its length, the full energy~(\ref%
{eq:19}) includes a contribution of the magnetized sheath. The
boundary condition~(\ref{eq:5a}) requires the continuity of the
normal component of magnetic induction $B_{n}^{\left( 1\right)
}=\mu _{0}\mu H_{n}^{\left( 2\right) }$. At the same time, the
total flux of magnetic induction is fixed by the flux quantization
in a superconductor. Accordingly, a typical magnetic induction
value in the magnet is $B\sim \Phi _{0}/\lambda ^{2}$, whereas the
magnetic field in the magnet $H\sim \Phi _{0}/\mu _{0}\mu \lambda
^{2}$\ is suppressed by the large $\mu $. Therefore, the
interaction energy proportional to $B\cdot H$\ is $\mu $\ times
reduced comparing with the vacuum case. Notice that this
conclusion is valid for the vortex loop nucleated at a SC/SM
interface of arbitrary form.

Now let us calculate the second part of the Gibbs
energy~(\ref{eq:39}). In the geometry of Fig.~1 the Meissner
current is perpendicular to the loop plane and almost constant in
the small loop region of size $a\ll \lambda $. In this case the
work of the Meissner current when the loop expands from the radius
$r=0$ to $a$\ reads simply as
\begin{equation}
\Delta W_{H}=\Phi _{0}\int\limits_{0}^{a}dr\int dl\
j_{M}=\frac{1}{2}\Phi _{0}\pi j_{s}a^{2},  \tag{42}  \label{eq:42}
\end{equation}

\noindent where $j_{s}$\ is a value of the screening current in a
place of the loop entry. The Meissner current in superconducting
cylinder has the only $z$\ component~\cite{VM}
\begin{equation}
j_{z}\left( \rho ,\varphi \right) =\frac{4\mu }{D}\frac{H_{0}}{\lambda }%
I_{1}\left( \rho /\lambda \right) \cos \varphi ,  \tag{43}
\label{eq:43}
\end{equation}%
\begin{equation}
D=\mu I_{0}\left( \frac{R_{1}}{\lambda }\right) \left[ \mu +1-\frac{\mu -1}{%
\left( 1+d/R_{1}\right) ^{2}}\right] -\frac{I_{1}\left(
R_{1}/\lambda \right) }{R_{1}/\lambda }\left( \mu ^{2}-1\right)
\frac{\left( d/R_{1}\right) \left( 2+d/R_{1}\right) }{\left(
1+d/R_{1}\right) ^{2}}, \tag{44}  \label{eq:44}
\end{equation}%
\noindent and the maximum magnitude of screening current
\begin{equation}
j_{s}=\frac{4\mu }{D}\frac{H_{0}}{\lambda }I_{1}\left(
R_{1}/\lambda \right) \tag{45}  \label{eq:45}
\end{equation}%
is achieved at the equatorial lines $\varphi =0$ and $\varphi =\pi
$\ where a vortex nucleates most probably.

The Gibbs energy of the vortex loop nucleating at $\varphi =0$\
renormalized by presence of the magnet,
\begin{equation}
G=\frac{\Phi _{0}^{2}}{4\pi \mu _{0}\lambda ^{2}}\pi a\ln \frac{a}{\xi }-%
\frac{1}{2}\Phi _{0}\pi j_{s}a^{2},  \tag{46}  \label{eq:46}
\end{equation}

\noindent grows with the radius $a$ from zero until it achieves a
maximum at some critical radius value $a_{m}$ defined by the
relation $\partial G/\partial a=0$. If the fluctuation vortex
reaches this size, further loop expansion becomes irreversible and
the vortex entry proceeds. Depending on a sample surface quality
vortex penetration may occur at different values of
the critical radius from the region $\xi <a_{m}<\lambda $ where the formula~(%
\ref{eq:46}) applies. The lower value corresponds to the case of
the ideal surface when the nucleation occurs at the scale of the
vortex core, $\xi $. The opposite limit describes a rough surface
with the typical imperfection size $\delta $ of the order of
$\lambda $ or larger. In general, a field of the first flux
penetration, $H_{p}$, defined by the condition $a_{m}=\min \left(
{\delta ,\lambda }\right) $, is given by
\begin{equation}
H_{p}=\frac{H_{p}^{0}}{4}\left\{ \frac{I_{0}\left( R_{1}/\lambda \right) }{%
I_{1}\left( R_{1}/\lambda \right) }\left[ \mu +1-\frac{\mu
-1}{\left( 1+d/R_{1}\right) ^{2}}\right] -\frac{\lambda }{\mu
R_{1}}\left( \mu ^{2}-1\right) \frac{\left( d/R_{1}\right) \left(
2+d/R_{1}\right) }{\left( 1+d/R_{1}\right) ^{2}}\right\} ,
\tag{47}  \label{eq:47}
\end{equation}%
\noindent where $H_{p}^{0}=\left( \Phi _{0}/4\pi \mu _{0}\lambda
a_{m}\right) \ln \left( ea_{m}/\xi \right) $ is a field of the
first flux penetration at a flat superconductor/vacuum boundary
adopting values between
the lower and the thermodynamic critical fields of the bulk materials~\cite%
{deGennes}.

\begin{figure}[h]
\includegraphics{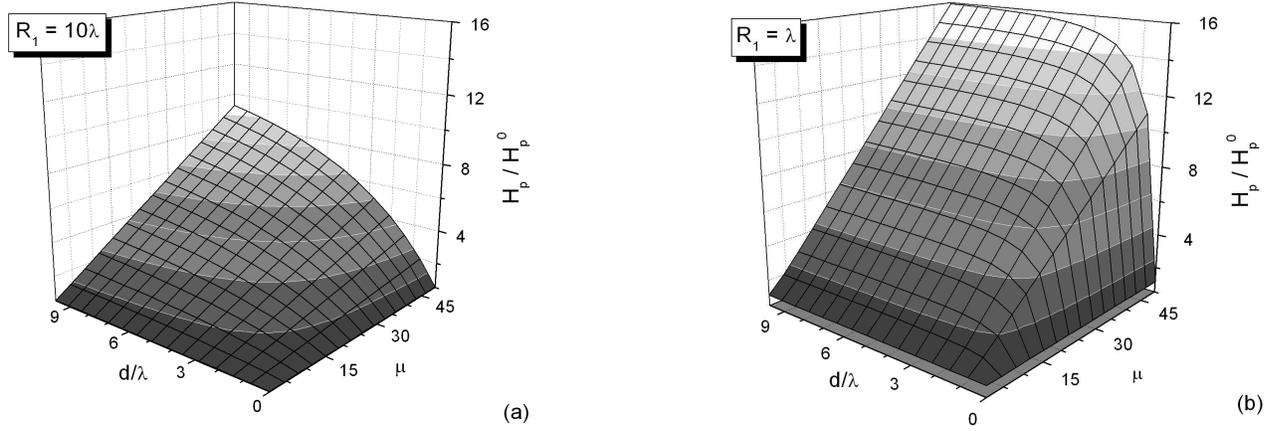}
\caption{The dependence of the field of the first flux penetration $H_{p}$
on the relative permeability $\protect\mu $ and on the thickness $d$ of the
magnet sheath for different values of the superconductor radius: a)~$R_{1}=10%
\protect\lambda $, and b)~$R_{1}=\protect\lambda $.}
\label{fig6}
\end{figure}

In Fig.~\ref{fig6} we present the dependence of the field $H_{p}$
on the relative permeability $\mu $ and the thickness $d$ for the
radii of superconductor $R_{1}=10\lambda $ and $R_{1}=\lambda $.
One can see that this dependence is monotonic for both cases of
thick and thin superconductor, contrary to the same dependences of
the transverse lower critical field $H_{c1}$, and differs only by
the scale of magnitudes. With increase of $\mu $ at fixed
thickness $d$\ the $H_{p}\left( \mu \right) $\ dependence
approximates to the linear one. A monotonic behavior is also
demonstrated by the $H_{p}\left( d\right) $\ dependence with
increase of $d$
at fixed permeability $\mu $. Considering practically interesting case $%
R_{1}\gg \lambda $ we write down
\begin{equation}
H_{p}=\frac{H_{p}^{0}}{4}\left[ \mu +1-\frac{\mu -1}{\left(
1+d/R_{1}\right) ^{2}}\right] .  \tag{48}  \label{eq:48}
\end{equation}%
In the opposite limit $R_{1}\ll \lambda $\ we obtain
\begin{equation}
H_{p}=H_{p}^{0}\frac{\lambda }{4\mu R_{1}}\left[ \left( \mu +1\right) ^{2}-%
\frac{\left( \mu -1\right) ^{2}}{\left( 1+d/R_{1}\right)
^{2}}\right] . \tag{49}  \label{eq:49}
\end{equation}

\subsection{Nucleation of vortex loop in presence of transport current}

Next, we consider the situation when the superconductor carries
also a transport current. The appearance of a total transport
current $J$ results
in an additional angle independent $z$-component of the current density~\cite%
{Schafroth,Clem}
\begin{equation}
j_{tr}\left( \rho \right) =\frac{J}{2\pi R_{1}\lambda
}\frac{I_{0}\left( \rho /\lambda \right) }{I_{1}\left(
R_{1}/\lambda \right) },  \tag{50} \label{eq:50}
\end{equation}

\noindent which is superimposed on the screening
current~(\ref{eq:43}). Similarly to the latter, the transport
current density remains constant within the loop and equal to the
surface value
\begin{equation}
j_{s,tr}=\frac{J}{2\pi R_{1}\lambda }\frac{I_{0}\left(
R_{1}/\lambda \right) }{I_{1}\left( R_{1}/\lambda \right) }.
\tag{51}  \label{eq:51}
\end{equation}

\noindent This surface magnitude should be simply added to the
maximum value
of the screening current~(\ref{eq:45}) and substituted in the Gibbs energy~(%
\ref{eq:46}). Using the criterion $a_{m}=\min \left( {\delta ,\lambda }%
\right) $ to define the critical current of the first flux penetration, $%
J_{c}$, we find the average density of the transport critical current, $%
j_{c}=J_{c}/\pi R^{2}$, for $H_{0}<H_{p}$
\begin{equation}
j_{c}(H_{0})=\frac{2I_{1}\left( R_{1}/\lambda \right)
}{R_{1}I_{0}\left( R_{1}/\lambda \right) }\left[
H_{p}^{0}-\frac{4\mu H_{0}}{D}I_{1}\left( R_{1}/\lambda \right)
\right] .  \tag{52}  \label{eq:52}
\end{equation}

\noindent In the limits of thin and thick superconductor core we
obtain, respectively,
\begin{equation}
j_{c}(H_{0})=\frac{H_{p}^{0}}{\lambda }-\frac{4\mu H_{0}R_{1}}{\lambda ^{2}}%
\left[ \left( \mu +1\right) ^{2}-\frac{\left( \mu -1\right)
^{2}}{\left( 1+d/R_{1}\right) ^{2}}\right] ^{-1},\qquad R_{1}\ll
\lambda ,  \tag{53} \label{eq:53}
\end{equation}

\begin{equation}
j_{c}(H_{0})=\frac{2H_{p}^{0}}{R_{1}}-\frac{8H_{0}}{R_{1}}\left[ \mu +1-%
\frac{\mu -1}{\left( 1+d/R_{1}\right) ^{2}}\right] ^{-1},\qquad
R_{1}\gg \lambda .  \tag{54}  \label{eq:54}
\end{equation}

\noindent One can see that in view of relatively large values of
the permeability in soft-magnet materials the field dependence of
$j_{c}$ remains linear up to the field $H_{0}\lesssim H_{p}$, that
corresponds to the experimental situation (see, for example,
Ref.~\cite{Dou1}).

For estimation at practically interesting temperature of
$32~\text{K}$ we take a reasonable $\mu \cong 50$ for $\text{Fe}$
and $d/R_{1}\cong 1/2$ from Refs.~\cite{Dou1,Dou2,Dou3} and
thermodynamic parameters of $\text{MgB}_{2} $ from
Ref.~\cite{Finnemore1} which gives possible values of $j_{c}(0) $
between $7.0\cdot 10^{3}$ and $4.4\cdot 10^{4}~\text{A/cm}^{2}$
and of $\mu
_{0}H_{p}$ between $0.16$ and $1.02~$T. The field dependence of $%
j_{c}(H_{0}) $ remains very weak up to the fields comparable with
$H_{p}$. In view of the relatively low values of critical
temperature,
Ginzburg-Landau parameter and anisotropy of the polycrystalline $\text{MgB}%
_{2}$ (Ref.~\cite{Finnemore1}) we assume that thermally activated
penetration through the surface barrier~\cite{Petukhov,Koshelev}
is negligible. These estimates are in a good agreement with the
results of Refs.~\cite{Dou1,Dou2,Dou3}.

\section{Conclusions}

To summarize, we have described the structure of an arbitrary
plane vortex in a type-II superconductor cylindrical filament
covered by a coaxial soft magnet sheath when it is exposed to an
external transverse magnetic field and carries a transport
current. We have derived general expressions for the magnetic
field components, self-energy and magnetic moment of the vortex.
Using these expressions, we have established that the self-energy
of the vortex lying along the sample diameter as well as that of
the small vortex loop nucleated at the interface between the
superconductor and the magnet is practically independent on the
parameters of magnet sheath, i.e. its permeability and thickness.
This paradoxical property is due to the phenomenon of flux
quantization in the superconductor.

We have found that the dependence of the magnetic moment of an
arbitrary plane vortex on both permeability and thickness of the
magnet sheath is not sensitive to the specific form of the vortex.
At the same time this dependence is qualitatively different for
thick and thin superconductors. For the samples with radius much
larger than $\lambda $ it reveals the minimum at some value of the
permeability for any fixed thickness of the sheath whereas for
thin superconductors having a radius less than $\lambda $ the
magnetic moment of vortex monotonously increases when the
permeability and/or sheath thickness increase. Such a behavior of
the magnetic moment causes the inverted dependence of the
transverse lower critical field, the exact expression of which was
derived. This field has a maximum at some value of the relative
permeability for any fixed thickness of the magnet sheath for the
case of the thick superconductor, while it decreases monotonously
with both permeability and magnet thickness increase for the case
of the thin superconductor.

We have considered also how the magnet sheath changes the
conditions for the flux penetration in the superconductor (the
Bean-Livingston barrier) when the sample is exposed to the
external transverse magnetic field and/or carries the transport
current. The expressions for the critical field and the critical
current of the first flux penetration have been derived. We found
that, due to presence of the magnet sheath, the critical field of
the first flux entry can be strongly enhanced. In contrast to the
lower critical field, it strongly increases in both thick and thin
superconductor cases, when the magnetic permeability of the sheath
is large. Obtained results have shown that, due to the magnet
sheath, the Meissner state in the superconductor filament can be
effectively preserved in wide region of the external magnetic
field (or transport current) magnitudes.

Notice also, that the above studied modification of the
Bean-Livingston
barrier is not reduced to the shielding effect of the magnetic sheath solely~%
\cite{Campbell2,DouAC1,DouAC2,DouAC3} because the latter does not
account for the magnetic flux expelling from the filament itself.
For example, in the limit $R_{1}\gg \lambda ,d$ the maximum field
on the superconductor surface amounts to $H_{max}=2H_{0}/(1+\mu
d/R_{1})$ while for the shielding
of the normal core it is $H_{max}=H_{0}/(1+\mu d/2R_{1})$ (see Ref.~\cite%
{Batygin}). The difference between these two values may be
substantial if the parameter $\mu d/R_{1}$\ is not large. The
observed range of external magnetic fields, where the critical
transport current remains virtually field-independent, exceeds
substantially the characteristic field of
effective shielding by the magnet sheath alone which was discussed in Ref.~%
\cite{Dou2}. On the other hand, estimations of the field for the
first flux entry, $H_{p}$, correlates well with the region of weak
field dependence of the critical current in MgB$_{2}$/Fe
cylindrical wires~\cite{Dou1,Dou2}. Reduced flux penetration in
the superconductor in this field region entails
naturally reduced AC losses which was also observed in Refs.~\cite%
{DouAC1,DouAC2,DouAC3}.

Finally, we have considered the simple model where the relative
permeability is a sole material characteristic of the magnet and
neglected its possible dependence on the applied field which may
be important~\cite{VM}. Nevertheless, obtained results are in good
qualitative and quantitative agreement with the existent
experiments and could be used to optimize the superconducting and
current-carrying parameters of SC/SM heterostructures.

\begin{acknowledgments}
We are grateful to H.~Rauh, H.~C.~Freyhardt, Ch.~Jooss, A.~V.~Pan,
J.~Horvat, S.~X.~Dou, M.~D.~Sumption, F.~M.~Peeters, A.~Gurevich
and V.~Vinokur for stimulating discussions. This work was
supported by a research grant of the German Research Foundation
(DFG). The support from the ESF VORTEX\ Program is also
acknowledged.
\end{acknowledgments}

\appendix

\section{The coefficients in Equations (15)-(18)}

The coefficients in Eqs.~(\ref{eq:15})-(\ref{eq:18}) read

\begin{equation}
\alpha _{k,m}=A_{1}~\frac{B_{+}+B_{-}-B_{3}}{\mu _{0}\Delta
},\quad \beta _{k,m}=A_{2}~\frac{B_{+}+B_{-}-B_{3}}{\mu _{0}\Delta
},  \label{A1}
\end{equation}

\begin{align}
F_{k,m}^{\pm }& =B_{\pm }-\alpha _{k,m}\mu _{0}\left\vert
k\right\vert \left[ \frac{I_{m\pm 1}\left( \left\vert k\right\vert
R_{1}\right) }{I_{m\pm 1}\left( QR_{1}\right) }+\left( \mu
-1\right) \frac{I_{m}^{\prime }\left(
\left\vert k\right\vert R_{1}\right) }{I_{m\pm 1}\left( QR_{1}\right) }%
\right]  \label{A2} \\ & -\beta _{k,m}\mu _{0}\left\vert
k\right\vert \left[ -\frac{K_{m\pm 1}\left( \left\vert
k\right\vert R_{1}\right) }{I_{m\pm 1}\left( QR_{1}\right)
}+\left( \mu -1\right) \frac{K_{m}^{\prime }\left( \left\vert
k\right\vert R_{1}\right) }{I_{m\pm 1}\left( QR_{1}\right)
}\right] ,  \notag
\end{align}

\begin{equation}
C_{k,m}=ik\mu _{0}\left[ \alpha _{k,m}~\frac{I_{m}\left(
\left\vert
k\right\vert R_{1}\right) }{I_{m}\left( QR_{1}\right) }+\beta _{k,m}~\frac{%
K_{m}\left( \left\vert k\right\vert R_{1}\right) }{I_{m}\left(
QR_{1}\right) }\right] ,  \label{A3}
\end{equation}

\begin{equation}
\Psi _{k,m}=\frac{1}{\left\vert k\right\vert R_{2}I_{m}\left(
\left\vert k\right\vert R_{2}\right) I_{m}^{\prime }\left(
\left\vert k\right\vert R_{2}\right)
}~\frac{B_{+}+B_{-}-B_{3}}{\mu _{0}\Delta },  \label{A4}
\end{equation}

\noindent where
\begin{equation}
A_{1}=\frac{\left( 1-\mu \right) }{\mu }~\frac{K_{m}\left(
\left\vert k\right\vert R_{2}\right) }{I_{m}\left( \left\vert
k\right\vert R_{2}\right)
}~\frac{K_{m}^{\prime }\left( \left\vert k\right\vert R_{2}\right) }{%
I_{m}^{\prime }\left( \left\vert k\right\vert R_{2}\right) },\quad A_{2}=%
\frac{K_{m}\left( \left\vert k\right\vert R_{2}\right)
}{I_{m}\left( \left\vert k\right\vert R_{2}\right) }-\frac{1}{\mu
}\frac{K_{m}^{\prime }\left( \left\vert k\right\vert R_{2}\right)
}{I_{m}^{\prime }\left( \left\vert k\right\vert R_{2}\right) },
\label{A5}
\end{equation}

\begin{align}
\Delta & =A_{1}\left\{ \left\vert k\right\vert \left[
\frac{I_{m+1}\left(
\left\vert k\right\vert R_{1}\right) }{I_{m+1}\left( QR_{1}\right) }+\frac{%
I_{m-1}\left( \left\vert k\right\vert R_{1}\right) }{I_{m-1}\left(
QR_{1}\right) }\right] -\frac{2k^{2}}{Q}\frac{I_{m}\left(
\left\vert k\right\vert R_{1}\right) }{I_{m}\left( QR_{1}\right)
}\right.  \label{A6} \\ & \left. +\left( \mu -1\right) \left\vert
k\right\vert I_{m}^{\prime }\left( \left\vert k\right\vert
R_{1}\right) \left[ \frac{1}{I_{m+1}\left( QR_{1}\right)
}+\frac{1}{I_{m-1}\left( QR_{1}\right) }\right] \right\} \notag \\
& +A_{2}\left\{ -\left\vert k\right\vert \left[
\frac{K_{m+1}\left(
\left\vert k\right\vert R_{1}\right) }{I_{m+1}\left( QR_{1}\right) }+\frac{%
K_{m-1}\left( \left\vert k\right\vert R_{1}\right) }{I_{m-1}\left(
QR_{1}\right) }\right] -\frac{2k^{2}}{Q}\frac{K_{m}\left(
\left\vert k\right\vert R_{1}\right) }{I_{m}\left( QR_{1}\right)
}\right.  \notag \\ & \left. +\left( \mu -1\right) \left\vert
k\right\vert K_{m}^{\prime }\left( \left\vert k\right\vert
R_{1}\right) \left[ \frac{1}{I_{m+1}\left( QR_{1}\right)
}+\frac{1}{I_{m-1}\left( QR_{1}\right) }\right] \right\} , \notag
\end{align}

\begin{equation}
B_{\pm }=\frac{K_{m\pm 1}\left( QR_{1}\right) }{I_{m\pm 1}\left(
QR_{1}\right) }\int\limits_{0}^{R_{1}}d\rho ~\rho \eta _{k,m}^{\pm
}\left(
\rho \right) I_{m\pm 1}\left( Q\rho \right) ,\quad B_{3}=\frac{2R_{1}}{%
\lambda ^{2}Q}K_{m}\left( QR_{1}\right) \Phi _{k,m}^{\rho }\left(
R_{1}\right) ,  \label{A7}
\end{equation}

\noindent the prime denotes the derivative of the Bessel function
with respect to its argument.

\section{The free energy of a hybrid SC/SM structure}

Let us calculate the excess free energy due to presence of
vortices in an arbitrary hybrid system composed of the
superconductor and insulating soft-magnet components with respect
to the energy of that system in the flux-free Meissner state.
Taking into consideration the potential nature of the magnetic
field outside the superconductor we write this energy, using the
London approximation, in the following way
\begin{equation}
F=\frac{1}{2\mu _{0}}\int dV^{\left( 1\right) }\left[
\mathbf{b}^{\left( 1\right) 2}+\lambda ^{2}\left(
\text{curl~}\mathbf{b}^{\left( 1\right)
}\right) ^{2}\right] +\frac{1}{2}\int dV^{\left( 2\right) }~\mathbf{B}%
^{\left( 2\right) }\mathbf{h}^{\left( 2\right) }+\frac{\mu
_{0}}{2}\int dV^{\left( 3\right) }\left( \mathbf{h}^{\left(
3\right) }\right) ^{2}, \label{B1}
\end{equation}

\noindent where $\mathbf{b}^{\left( 1\right) }$\ is the magnetic
field of
vortices in the superconductor, $\mathbf{h}^{\left( 2\right) }$ and $\mathbf{%
h}^{\left( 3\right) }$ are the magnetic fields in the magnet media
and in the surrounding free space, respectively,
$\mathbf{B}^{\left( 2\right) }=\mu _{0}\mu \mathbf{h}^{\left(
2\right) }$ is the magnetic induction in the
magnet media. With the vector identity div$\left( \mathbf{a}\times \mathbf{c}%
\right) =\mathbf{c}~$curl$~\mathbf{a-a}~$curl$~\mathbf{c}$ and
with the London equation~(\ref{eq:6}) in the superconducting
region the first term of Eq.~(\ref{B1}) becomes
\begin{eqnarray}
F^{\left( 1\right) } &=&\frac{1}{2\mu _{0}}\int dV^{\left(
1\right) }\left\{
\mathbf{b}^{\left( 1\right) 2}+\lambda ^{2}\left[ \text{curl~curl~}\mathbf{b}%
^{\left( 1\right) }+\text{div}\left( \mathbf{b}^{\left( 1\right)
}\times \text{curl~}\mathbf{b}^{\left( 1\right) }\right) \right]
\right\}  \label{B2}
\\
&=&\frac{1}{2\mu _{0}}\int dV^{\left( 1\right) }\
\mathbf{b}^{\left( 1\right) }\mathbf{\Phi +}\frac{\lambda
^{2}}{2\mu _{0}}\int dS^{\left( 1\right) }\mathbf{n}^{\left(
1\right) }\left. \left( \mathbf{b}^{\left( 1\right) }\times
\text{curl~}\mathbf{b}^{\left( 1\right) }\right) \right\vert
_{S^{\left( 1\right) }},  \notag
\end{eqnarray}

\noindent The second and third terms of Eq.~(\ref{B1}) are
transformed, with the definition $\mathbf{h}^{\left( 2,3\right)
}=-\mathbf{\nabla }\psi
^{\left( 2,3\right) }$ and with the identity div$\left( \psi \mathbf{a}%
\right) =\psi $div$~\mathbf{a+a\nabla }\psi $, in the following
surface integrals
\begin{eqnarray}
F^{\left( 2\right) } &=&-\frac{1}{2}\int dV^{\left( 2\right) }\left[ \text{%
div}\left( \psi ^{\left( 2\right) }\mathbf{B}^{\left( 2\right)
}\right) -\psi ^{\left( 2\right) }\text{div~}\mathbf{B}^{\left(
2\right) }\right] \label{B3} \\ &=&-\frac{1}{2}\int dS^{\left(
2\right) }\mathbf{n}^{\left( 2\right) }\left. \left( \psi ^{\left(
2\right) }\mathbf{B}^{\left( 2\right) }\right) \right\vert
_{S^{\left( 2\right) }},  \notag
\end{eqnarray}%
\begin{equation}
F^{\left( 3\right) }=-\frac{\mu _{0}}{2}\int dV^{\left( 3\right) }\mathbf{h}%
^{\left( 3\right) }\mathbf{\nabla }\psi ^{\left( 3\right) }=-\frac{\mu _{0}}{%
2}\int dS^{\left( 3\right) }\mathbf{n}^{\left( 3\right) }\left.
\left( \psi ^{\left( 3\right) }\mathbf{h}^{\left( 3\right)
}\right) \right\vert _{S^{\left( 3\right) }}.  \label{B4}
\end{equation}

\noindent In Eqs.$~$(\ref{B2}-\ref{B4}) $\mathbf{n}^{\left(
j\right) }$ denotes the outer normal to the surface~$S^{\left(
j\right) }$ of the corresponding region. Next we assume that the
field $\mathbf{h}^{\left( 3\right) }$\ decreases sufficiently
fast, so that the part of surface
integral~(\ref{B4}) corresponding to the integration at $\mathbf{r}%
\rightarrow \infty $ vanishes. The rest of this surface integral,
i.e. integral over the magnet outer surface, is compensated by the
corresponding part of the surface integral~(\ref{B3}) by virtue of
the boundary condition~(5b) and we have
\begin{eqnarray}
F &=&\frac{1}{2\mu _{0}}\int dV^{\left( 1\right) }\
\mathbf{b}^{\left( 1\right) }\mathbf{\Phi +}\frac{\lambda
^{2}}{2\mu _{0}}\int dS^{\left( 1\right) }\mathbf{n}^{\left(
1\right) }\left. \left( \mathbf{b}^{\left( 1\right) }\times
\text{curl~}\mathbf{b}^{\left( 1\right) }\right) \right\vert
_{S^{\left( 1\right) }}  \label{B5} \\ &&+\frac{1}{2}\int
dS^{\left( 1\right) }\mathbf{n}^{\left( 1\right) }\left. \left(
\psi ^{\left( 2\right) }\mathbf{B}^{\left( 2\right) }\right)
\right\vert _{S^{\left( 1\right) }}.  \notag
\end{eqnarray}

\noindent Taking into account that the second term of
Eq.~(\ref{B5}) contains only tangential components of
$\mathbf{b}^{\left( 1\right) }$, we transform this integral in the
following way:
\begin{eqnarray}
&&-\int dS^{\left( 1\right) }\mathbf{n}^{\left( 1\right) }\left.
\left(
\mathbf{\nabla }\psi ^{\left( 2\right) }\times \text{curl~}\mathbf{b}%
^{\left( 1\right) }\right) \right\vert _{S^{\left( 1\right) }}
\label{B6} \\ &=&\int dS^{\left( 1\right) }\mathbf{n}^{\left(
1\right) }\left. \left[ \psi
^{\left( 2\right) }\times \text{curl~curl~}\mathbf{b}^{\left( 1\right) }-%
\text{curl~}\left( \psi ^{\left( 2\right)
}\text{curl~}\mathbf{b}^{\left( 1\right) }\right) \right]
\right\vert _{S^{\left( 1\right) }}.  \notag
\end{eqnarray}

\noindent It is easy to show that, by applying the Stokes theorem,
the last integral in Eq.~(\ref{B6}) vanishes (see, for example,
Ref.~\cite{Kogan}). Finally, applying the boundary condition~(5a)
to the normal components of the induction we obtain
\begin{equation}
F=\frac{1}{2\mu _{0}}\int dV^{\left( 1\right) }\
\mathbf{b}^{\left( 1\right) }\mathbf{\Phi +}\frac{1}{2}\int
dS^{\left( 1\right) }\left. \left( \psi ^{\left( 2\right)
}\mathbf{\Phi n}^{\left( 1\right) }\right) \right\vert _{S^{\left(
1\right) }}.  \label{B7}
\end{equation}

\noindent Let us note that Eq.~(\ref{B7}) does not contain
explicitly any characteristics of the magnetic media which enter
only the expressions for quantities $\mathbf{b}^{\left( 1\right)
}$\ and $\psi ^{\left( 2\right) }$. From Eq.~(\ref{B7}), the
formula~(\ref{eq:19}) immediately follows for the geometry
considered in the paper.

The obtained general expression~(\ref{B7}) for the free energy is
applicable for any configuration of superconductor and magnet
components (for example, for multifilamentary
superconductor/magnet wires or tapes) and is valid in the whole
region of the vortex state in a type-II superconductor.

\bibliographystyle{plain}
\bibliography{apssamp}

\end{document}